\documentstyle[11pt,epsfig]{article} 
\def\baselinestretch{1.3}
\newcommand{\ba}{\begin{array}}
\newcommand{\ea}{\end{array}}
\newcommand{\bd}{\begin{displaymath}}
\newcommand{\ed}{\end{displaymath}}
\newcommand{\be}{\begin{equation}}
\newcommand{\ee}{\end{equation}}
\newcommand{\bea}{\begin{eqnarray}}
\newcommand{\eea}{\end{eqnarray}}

\def\q2 {q^2}

\begin{document}
\begin{flushright}
{\large DESY-98-134\\ 
\large MRI-PHY/P980857 \\
August, 1998  \\ 
hep-ph/9808265}
\end{flushright}

\begin{center}
{\Large\bf
Correlation between neutrino oscillations and collider signals of 
supersymmetry in an $R$-parity violating model}\\[20mm]
Biswarup Mukhopadhyaya and Sourov Roy\\
{\em Mehta 
Research Institute,\\
Chhatnag Road, Jhusi, Allahabad - 211 019, India}, 
\\[10mm]
Francesco Vissani\\{\em Theory Group, DESY\\  Notkestrasse 85,
D - 22603 Hamburg, Germany}
\\[20mm] 
\end{center}
\begin{abstract}
Motivated by the recent Super-Kamiokande results on atmospheric neutrinos, 
we incorporate massive neutrinos, 
with large angle 
oscillation between the second and third generations, in a scenario
with $R$-parity violating supersymmetry. We emphasize 
the testability of such
models through the observation of comparable numbers of muons and 
taus, produced together with the W-boson, in decays
of the lightest neutralino. A distinctly measurable decay gap is another
remarkable feature of such a scenario.  

\end{abstract}

\vskip 1 true cm

\newpage
\setcounter{footnote}{0}

\def\baselinestretch{1.8}
~~The recent results from the Super-Kamiokande (SK) experiment on 
atmospheric neutrinos
strengthen the already existing evidence in favour 
of neutrino oscillations \cite{1}. 
The zenith angle distribution of the events  
can be explained assuming that
$\nu_{\mu}$ oscillates dominantly 
into $\nu_{\tau},$ or possibly into 
a sterile neutrino.
A large mixing angle is strongly suggested by the data, 
with a best-fit mass-squared difference 
($\Delta m^2$) of $2.2 \times 10^{-3}\ {\rm {eV^2}}$. 
The Soudan-2 results support these
claims, their favoured $\Delta m^2$ 
being above 
$1 \times 10^{-3}\ {\rm {eV^2}}$ \cite{3}.  
Further evidence comes from 
MACRO data on upward 
going muons \cite{3.5}. 
Furthermore, oscillations of the above kind allow one to  
account for the solar neutrino 
flux deficit as well\cite{2}.

The very existence of mass and mixing 
in the neutrino sector takes one beyond 
the standard model, and the obvious question that arises is about a 
theoretical framework that naturally accommodates the observed data, 
particularly the large mixing angle. 
Several classes of models for neutrino 
masses exist in the literature, a few examples being
see-saw models with massive right-handed neutrinos \cite{4},
models
with radiative generation of neutrino 
masses \cite{5}, or Majorana masses 
induced by a Higgs triplet \cite{6}. 
However, it requires considerable 
manoeuvring \cite{7} in each of the 
above cases in order to achieve 
large angle oscillations 
while keeping everything else 
consistent from a phenomenological point of view.

An interesting  possibility is 
supersymmetry (SUSY) \cite{8} where non-zero 
neutrino masses can be envisioned 
once we allow the violation of $R$-parity, 
defined as $R = (-1)^{3B + L + 2S}$ 
(where the baryon and 
lepton numbers $B$ and $L$ are assigned
to the supermultiplets).
This is in perfect consistence with all 
experimental observations, 
including proton stability, 
if either baryon or lepton 
number is still conserved. 
The standard way of writing down the
$R$-parity violating effects \cite{9} 
is to consider the following additional 
terms in the MSSM superpotential(suppressing ${\rm SU(2)}$ indices) :

\begin{equation}
W_{\not R} = \lambda_{ijk} L_i L_j E_k^c +
\lambda_{ijk}' L_i Q_j D_k^c + 
\lambda_{ijk}''U_i^c D_j^c D_k^c + \epsilon_i L_i H_2
\end{equation}

\noindent
where, in a phenomenologically 
viable scenario, it is safe to either keep only
the $\lambda''$-term or drop it. 
The $\lambda$ and $\lambda'$-terms give
rise to neutrino masses at the one-loop level; the bilinear
terms $\epsilon_{i}L_{i}H_2$ in turn
(in association with non-vanishing sneutrino vacuum
expectation values) allow one to augment the neutralino mass matrix 
with the neutrinos, leading to a 
see-saw type mass for the latter. A number 
of studies on both possibilities have already been published 
\cite{10,11,12,13}.

In this note, we are mainly 
concerned with the latter scenario 
for massive neutrinos, in the light
of the current experimental indications  on
atmospheric neutrinos. We point out a logical way of 
understanding the observed results, which
still keeps enough room for
a solution of the solar neutrino puzzle. 
In doing this, we neither have to  assume a near-equality of
tree-and loop-induced masses \cite{14}, 
nor do we use only loop-induced 
effects \cite{15} which run the risk of predicting an excess of 
flavour-changing neutral current (FCNC) phenomena. In our formulation, 
all extra parameters over and above the ones of MSSM can be traded by the 
neutrino mass-squared difference required by the SK data and the large 
angle of oscillation.  And, most importantly, 
we predict that in a scenario 
like this, the decay of the lightest neutralino $\tilde \chi^0_1$
(which is unstable once 
$R$-parity is violated) will produce comparable 
numbers of muons and taus
as a result of large-angle mixing. 
This provides one with a method to
test (or falsify) such models of 
neutrino mass generation in collider
experiments.  
                           
We start by assuming bilinear 
terms in the superpotential (1) for
$i = 2,3$ only. 
Next, we rotate away the terms proportional to
$\epsilon_2$ and $\epsilon_3$ by 
redefining the lepton and Higgs superfields.
However, this does not eliminate the 
impact of the bilinear terms, since such 
rotation has its effect on the scalar potential \cite{16}, causing the 
sneutrinos to have non-vanishing 
vacuum expectation values (vev) in general. 
These vev's (denoted here by $v_2$ and $v_3$) 
induce terms in the neutralino 
mass matrix via the 
neutrino-sneutrino-Bino (or ${W_3}$-ino) 
interactions. Thus in this basis (which, 
for our purpose, serves as the flavour basis) the 
$6 \times 6$ neutralino mass matrix becomes 

\begin{equation}
{\cal M} =  \left( \begin{array}{cccccc}
  0 & -\mu & \frac {gv} {\sqrt{2}} & 
  -\frac {g'v} {\sqrt{2}} & 0 & 0 \\
  -\mu & 0 & -\frac {gv'} {\sqrt{2}} 
       & \frac {g'v'} {\sqrt{2}} & 0 & 0 \\
 \frac {gv} {\sqrt{2}} & -\frac {gv'} {\sqrt{2}} & M & 0 & -\frac {gv_3} 
 {\sqrt{2}} & -\frac {gv_2} {\sqrt{2}} \\
 -\frac {g'v} {\sqrt{2}} & \frac {g'v'} {\sqrt{2}} & 0 & M' & 
  \frac {g'v_3} {\sqrt {2}} & \frac {g'v_2} {\sqrt {2}} \\
 0 & 0 & -\frac {gv_3} {\sqrt {2}} & \frac {g'v_3} {\sqrt {2}} & 
 0 & 0  \\
 0 & 0 & -\frac {gv_2} {\sqrt {2}} & \frac {g'v_2} {\sqrt {2}} & 
 0 & 0 
 \end{array}  \right)    
\end{equation}                                 

\noindent
where the successive rows and columns correspond to
(${\tilde H}_2, {\tilde H}_1, -i\tilde{W_3}, 
-i\tilde{B}, \nu_\tau, \nu_\mu$).
Here 

$$
v\ \ (v') = \sqrt{2}\ {\left(\frac {m^2_Z} {\bar{g}^2} 
 - \frac {v^2_2+v_3^2} {2} \right)}
^{\frac {1} {2}} {{\sin} \beta}\ \ ({{\cos} \beta})
$$

\noindent
$M$ and $M'$ are the ${\rm SU(2)}$ and ${\rm U(1)}$ 
gaugino mass parameters 
respectively, $\mu$, 
the Higgsino mass parameter,
$m_Z$, the $Z$ boson mass, 
and $\bar{g}=\sqrt{g^2+{g'}^2}.$ 
One can define two states $\nu_3$ and $\nu_2$, where

\begin{equation}
\nu_3 =\cos \theta\ \nu_\tau  + \sin \theta\ \nu_\mu\cdot 
\end{equation}

\noindent
and $\nu_2$ is the 
orthogonal combination, 
the neutrino mixing angle being
\begin{equation}
\cos \theta = \frac{v_3}{\sqrt{v^2_2 + v^2_3}},\ \ \ \ \
\sin \theta = \frac{v_2}{\sqrt{v^2_2 + v^2_3}}
\end{equation}                  
Clearly, the state $\nu_2$ remains 
massless, whereas $\nu_3$ acquires a
see-saw type mass:

\begin{equation}
m_{\nu_3}\approx 
-\frac{\bar{g}^2 (v_2^2+v_3^2)}{2\ \bar{M}}\times
\frac{\bar{M}^2}{M M' - m_Z^2\ \bar{M}/\mu \ \sin 2\beta }
\end{equation}
where we introduced 
$\bar{g}^2 \bar{M}=g^2 M'+{g'}^2 M.$ 
The first term is very similar to 
the usual see-saw formula, with the 
only difference that couplings between the light and
the heavy states is in the present case 
due to gauge interactions. 

The massive state $\nu_3$ 
can be naturally used to account for
atmospheric neutrino oscillations,
with $\Delta m^2 = m_{\nu_3}^2.$
Large angle mixing between the $\nu_\mu$ and 
the $\nu_\tau$ corresponds to the 
situation where $v_2 \simeq v_3$.
Before we discuss how 
this assumption 
affects the observable signatures for
supersymmetry at colliders,
two remarks are in order: \\
1) The reason why only one neutrino becomes massive 
is that only one ``heavy'' state, 
the Zino, is coupled to the neutrinos
(it corresponds to a see-saw formula 
in which only one right-handed neutrino has
Yukawa couplings with the left-handed neutrinos).\\
2) The formalism can be extended to include a 
subdominant component $\nu_e$ of 
the massive neutrino state $\nu_3$,
simply letting the vev $v_1$ to be non-zero
(future experimental data and 
analyses on neutrinos will permit us 
to assess the size of this component). 
 
Let us now turn to the phenomenological 
implication of the above scenario in 
the neutralino sector. 
In most models, the lightest neutralino is the lightest 
supersymmetric particle (LSP). 
The non-conservation of $R$-parity implies that
it can decay into particles with $R = +1$. 
Here, an interesting possibility 
arises exclusively from the bilinear $R$-violating terms \cite{17,18}. 
As a result of the mixing between 
neutrinos and neutralinos (as also between 
charged leptons and charginos) 
the LSP has the additional decay channels

\begin{equation}
\tilde \chi^0_1 \longrightarrow \nu_l Z \ (Z^*)\ \ \ \ \ l=e,\mu,\tau
\end{equation}

\noindent
and

\begin{equation}
\tilde \chi^0_1 \longrightarrow l W(W^*)\ \ \ \ \ l=e,\mu,\tau
\label{ccmode}
\end{equation}

\noindent
which are absent
with only the commonly discussed $\lambda$-and $\lambda'$-
terms at leading order.  
However, if the neutralino is lighter than the $W$ 
boson, then the 
resulting three-body decays give 
rise to final states which can also 
be produced by the trilinear $R$-violating 
interactions.  Thus, the signals for 
bilinear interactions are most prominent 
for $m_{\tilde \chi^0_1} > m_Z (m_W)$, 
which corresponds to a large part of the 
SUSY parameter space still allowed by experiments.

In the above range of neutralino masses, large angle mixing between
the second and third neutrino generations implies that $l$ in 
Eqn.(\ref{ccmode}) above
can be the muon or the 
tau with comparable probabilities. Thus one should see
muons and tau's in near-equal numbers, along with the 
$W$-boson, in the collider 
signals of the lightest neutralino if 
$R$-parity violating SUSY has to provide 
the mechanism of generating neutrino masses. 

In Fig.(1), we plot the branching ratios of the two-body decays  
as  functions of the neutralino mass. We demonstrate our point by
assuming maximal mixing, i.e.\ $\theta = \pi/4$, which is achieved by
setting $v_2 = v_3.$ 
In our calculation, the MSSM parameters
$\mu$, $\tan \beta$ and the universal 
gaugino mass $\tilde{M}$ are used as the input 
parameters. 
The gaugino mass
parameters $M$ and $M'$ have been 
assumed to be related 
by the condition
of (SU(5)) gaugino 
mass unification. 

\begin{figure}[hbt]
\centerline{\epsfig{file=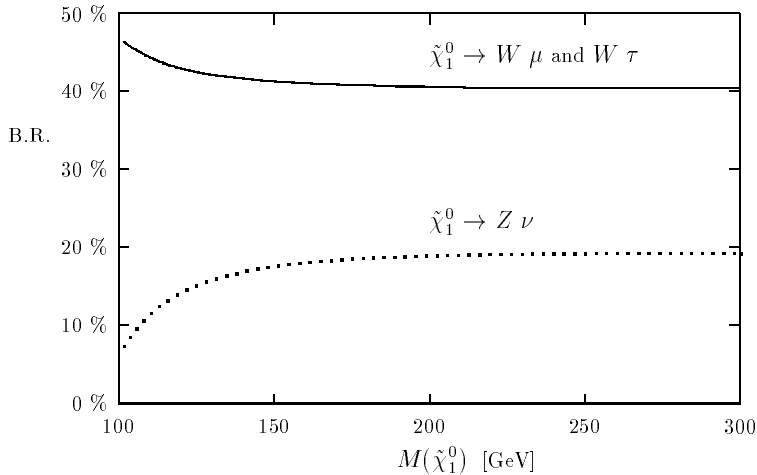,width=10cm}}
\caption{Branching ratio for
the decay channels of the lightest
supersymmetric state $\tilde \chi_1^0.$
The neutrino mixing angle is maximal, 
and $\Delta m^2=2.15 \times 10^{-3}\ {\rm {eV^2}}.$
The remaining supersymmetric parameters are chosen as:
$\mu=-500$ GeV, $\tan\beta=5.$}
\end{figure}

In our plot, both 
$l^{+} W^{-}$ and $l^{-}W^{+}$ ($l = \mu, \tau$) have 
been included in the charged current decay modes. We have checked that 
the relative strengths of the $W$-and $Z$-modes 
(which flatten out once the 
phase space suppressions become insignificant) remain roughly the same 
and insensitive to $\mu$ and 
${\rm tan} \beta$ so long as the LSP continues 
to be dominated by the Bino.  
The ratio between the two types of decays can 
change when there is a substantial 
Higgsino content of the LSP, which can 
consequently have couplings with the $W$ and the $Z$. 
Also, here we have 
neglected the two-body decay of 
the LSP with the lightest Higgs boson in 
the final state.   
The latter can be appreciable (up to about 25\% \cite{18}) 
when the LSP has a large Higgsino component, 
or when it is Bino-dominated 
and ${\rm tan} \beta$ is close to 1. 
However, none of the above situations 
alter the fact that  muons and taus 
are produced in comparable intensities 
as a result of neutralino decay, so 
long as the neutralino is massive enough 
for the two-body decays to be  allowed. 
Therefore, our main prediction 
continues to hold over the entire 
$\mu - \tan\beta$ space consistent with 
experiments.

Another useful test for this scenario can be performed by measuring the 
decay length of the lightest neutralino. This is given by the formula

\begin{equation}
L =  \frac{\hbar}{\Gamma} 
\times 
\frac{p}{M(\tilde{\chi}^0_1)} 
\end{equation}  

\noindent
where $\Gamma$ is the 
decay width of the 
lightest neutralino and
$p$ its momentum. In Fig.\ 2 we present a plot of the    
decay length against the neutralino mass for three different values of 
$\Delta m^2$, corresponding to the extreme limits allowed by SK data
for $\nu_\mu \rightarrow \nu_\tau$ oscillation. The decaying neutralino
is assumed to have an energy of 250 GeV.                   
As is expected, the decay 
length decreases for higher neutrino masses,
as a result of the enhanced probability of the flip between 
Bino and neutrinos, 
when the LSP is dominated by the Bino. 
What is interesting, however, is the fact that the decay lengths are
as large as about 0.1 - 10 millimeters even 
for the largest possible neutrino 
mass. This gives us an additional and 
rather interesting characterization
of the reaction $\tilde \chi^0 
\longrightarrow \tau(\mu) W$ at colliders,
in the assumption that these states are 
sufficiently light to be produced in next
colliders.

\begin{figure}[hbt]
\centerline{\epsfig{file=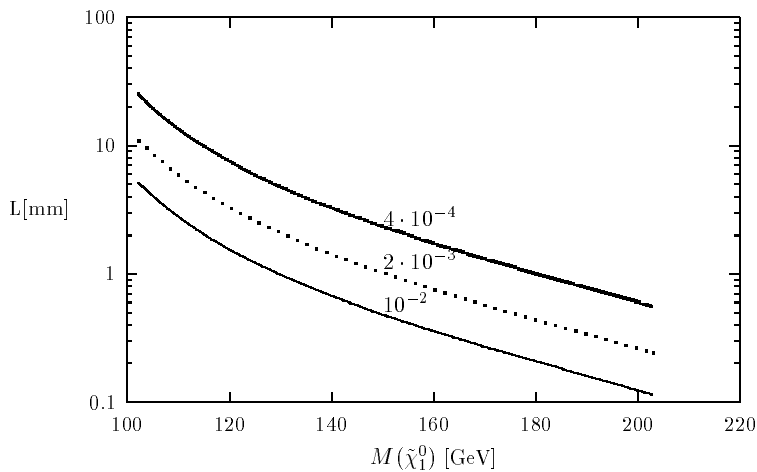,width=10cm}}
\caption{Decay length for $\tilde \chi_1^0$ for
three values of $\Delta m^2,$ indicated at the curves.
$\mu=-500$ GeV, $\tan\beta=5.$}
\end{figure}

Finally we consider 
the question of what effects do the trilinear $R$-violating
couplings $\lambda, \lambda'$ have on our scenario. 
Their  presence will
in general give rise to mass terms, both diagonal and off-diagonal,
generated at one-loop level involving all the neutrinos. 
The general expression
for these masses is

\begin{equation}
(m^{loop}_\nu)_{ij} \simeq \frac {3} {8\pi^2}  m^d_k m^d_p M_{SUSY} 
\frac {1} {m^2_{\tilde q}} {\lambda_{ikp}'\lambda_{jpk}'}  + 
\frac {1} {8\pi^2}  m^l_k m^l_p M_{SUSY} 
\frac {1} {m^2_{\tilde l}} {\lambda_{ikp}\lambda_{jpk}}  
\end{equation}         

\noindent 
where $m^{d,l}$ denote the down-type quark and charged lepton masses, 
respectively. 
${m^2_{\tilde l}}$, ${m^2_{\tilde q}}$ are the slepton and 
squark mass squared. $M_{SUSY}(\sim \mu)$ is the effective scale of 
supersymmetry breaking. 
If we want the mass thus induced for the second generation neutrino  
to be the right one to solve the solar neutrino problem, then one obtains  
some constraint on the value of the $\lambda'$s as well as $\lambda$s. 
In order to generate a splitting between the two 
residual massless neutrinos,
$\delta m^2 \simeq 5 \times 10^{-6}\ {\rm {eV^2}}$ 
(which is suggested 
for an MSW solution \cite{19}), a SUSY breaking 
mass of about 500 GeV implies 
$\lambda'\ (\lambda) \sim  10^{-4} - 10^{-5}$. 
Such a value of $\lambda'\ (\lambda)$ makes 
the three-body decays of the 
neutralino too small compared 
to the two-body decays when 
they are allowed. 
The mass-squared difference required for 
a vacuum oscillation solution \cite{20} 
to the solar puzzle requires even 
smaller values of $\lambda'(\lambda).$  
Thus the simultaneous presence of trilinear and bilinear R-violating
couplings are not expected to cause any
noticeable change in the muon 
vs.\ tau branching ratios 
in LSP decays that we are 
concerned with.

In conclusion, we have discussed 
an $R$-parity violating scenario which
can accommodate the large 
mixing angle suggested by the 
atmospheric neutrino anomaly.
The  signature of this scenario is the
production of comparable 
numbers of muons and tau's 
in the decay of the lightest neutralino
at colliders. 
In addition, the decay could lead to a
measurable secondary vertex (decay gap).
This provides one 
with the prospect of a verification  in collider 
experiments as to whether SUSY indeed 
is responsible for the masses and mixing
of neutrinos.

\noindent
{\bf Acknowledgement:} B.M.\ wishes to acknowledge the hospitality of
Peter Zerwas and the DESY Theory Group where this work was initiated.

\end{document}